# Rate of Channel Hardening of Antenna Selection Diversity Schemes and Its Implication on Scheduling

Dongwoon Bai, Patrick Mitran, Saeed S. Ghassemzadeh, Robert R. Miller, and Vahid Tarokh

**Abstract**

For a multiple antenna system, we compute the asymptotic distribution of antenna selection gain when the transmitter selects the transmit antenna with the strongest channel. We use this to asymptotically estimate the underlying channel capacity distributions, and demonstrate that unlike multiple-input/multiple-output (MIMO) systems, the channel for antenna selection systems hardens at a slower rate, and thus a significant multiuser scheduling gain can exist - $O(1/\log m)$ for channel selection as opposed to $O(1/\sqrt{m})$ for MIMO, where $m$ is the number of transmit antennas. Additionally, even without this scheduling gain, it is demonstrated that transmit antenna selection systems outperform open loop MIMO systems in low signal-to-interference-plus-noise ratio (SINR) regimes, particularly for a small number of receive antennas. This may have some implications on wireless system design, because most of the users in modern wireless systems have low SINRs.

## I. Introduction

The use of multiple transmit antennas has been studied for wireless links because of its promise of high spectral efficiency. When the receiver has full channel state information (CSI), the capacity of a MIMO channel is typically calculated under the assumption that either the transmitter has full CSI (closed loop MIMO) or no CSI (open loop MIMO) [1]. In both cases, in order to achieve data rates close to capacity, the implementation of various signal processing and RF units is needed. The underlying costs

Dongwoon Bai, Patrick Mitran, and Vahid Tarokh are with Harvard School of Engineering and Applied Sciences, Cambridge, MA 02138 USA (email: dbai@fas.harvard.edu; mitran@deas.harvard.edu; vahid@deas.harvard.edu).

Saeed S. Ghassemzadeh and Robert R. Miller are with AT&T Labs - Shannon Laboratory, Florham Park, NJ 07932 USA (email: saeedg@att.com; rrm@att.com).

may increase as the number of antennas increases. For some applications, this cost is prohibitive and motivates the studies of alternative antenna technologies.

Antenna selection schemes are attractive, since they can reduce hardware costs dramatically [2]–[7]. Some proposals consider the selection of more than one antenna and require more than one transmit chain. Nonetheless, in this paper, we are mainly interested in a transmit antenna selection scheme that selects the best channel between transmit antennas. We will demonstrate that under this scheme, there is no channel hardening, and thus significant multiuser scheduling gain can exist. This is unlike MIMO systems [8], where the asymptotic scheduling gain is zero and there is significant channel hardening.

To this end, we compute the asymptotic distribution of the selection gain and use this to asymptotically estimate the underlying channel capacity distributions. We note that the exact distribution of the selection gain has been computed in the literature and the channel capacity has been numerically and explicitly (as a series expansion) calculated. However, these exact values are not insightful in predicting if channel hardening exists or not, let alone the rate of hardening. For this purpose, we will invoke the theory of extreme order statistics assuming Rayleigh fading channels.

The outline of the paper is given next. In Section II, we present our system model. In Section III, we calculate the asymptotic distribution of selection gains and outage capacity gains. In Section IV, we obtain upper and lower bounds for the ergodic capacity. These results demonstrate that channel hardening occurs at much slower rate than MIMO, and thus significant multiuser scheduling gain can exist. Additionally, we will show that even without this scheduling gain, transmit antenna selection can outperform an open loop MIMO system in the low SINR regime for small number of receive antennas. In Section V we compute the scheduling gain. Finally, in Section VI, we present our conclusions and final comments while most of the proofs may be found in the Appendix.

## II. THE SYSTEM MODEL

We consider an $n \times m$ MIMO channel model, with $m$ transmit and $n$ receive antennas. The input-output relation is given by

$$\boldsymbol{y} = \boldsymbol{H}\boldsymbol{s} + \boldsymbol{w}. \tag{1}$$

The matrix $\boldsymbol{H}$ represents the channel matrix, and is assumed to be known at the receiver. The $m \times 1$ complex vector $\boldsymbol{s}$ is the transmitted signal vector, the $n \times 1$ vector $\boldsymbol{y}$ represents the received signal, and $\boldsymbol{w}$ is an $n \times 1$ zero-mean i.i.d. circularly symmetric complex Gaussian noise vector, with covariance matrix $\mathbb{E}[\boldsymbol{w}\boldsymbol{w}^\dagger] = \boldsymbol{I}_n$. Interference, if any, is assumed to be absorbed in $\boldsymbol{w}$. An average transmit power constraint $\mathbb{E}[\boldsymbol{s}^\dagger \boldsymbol{s}] \leq \rho$, is assumed, where $\rho \geq 0$.

Assuming that a transmit antenna selection system chooses to transmit only on the $l$-th antenna ($l \in \{1, \cdots, m\}$), the capacity is given by

$$C(\boldsymbol{H}, l) = \log_2 \left(1 + \rho \sum_{i=1}^{n} |H_{il}|^2\right). \tag{2}$$

Let $X_l = \sum_{i=1}^{n} |H_{il}|^2$, then the best selection strategy is to choose antenna

$$l^* = \arg\max_l [\log_2(1 + \rho X_l)] = \arg\max_l X_l \tag{3}$$

for transmission. This scheme is of interest, since it eliminates the need to feed back the channel matrix $\boldsymbol{H}$. In fact, the receiver needs only to feed back the index of the best transmit antenna to the transmitter, requiring only $\log_2 m$ bits of feedback information.

We are interested to see if there is asymptotic channel hardening for such a system, i.e. whether or not the underlying scheduling gain asymptotically goes to zero and the rate of which this occurs. To this end, we assume that $\boldsymbol{H}$ has independent zero-mean complex Gaussian entries with variance 1/2 per real components. This is a flat fading channel model. Thus $X_l$ has chi-square distribution with $2n$ degrees of freedom and $X_l$ is independent of $X_{l'}$ whenever $l \neq l'$.

For any set of i.i.d. random variables $Z_1, ..., Z_m$, we use the notation $Z_{(m)}$ to denote $\max_{1 \leq i \leq m} Z_i$. Using this notation, the expected received SINR for the transmit antenna selection strategy is given by $\rho \cdot X_{(m)}$. The asymptotic distribution of $X_{(m)}$ for large $m$ is of interest, since the ergodic capacity of the selection scheme with the optimal choice of transmit antenna, is given by

$$\begin{aligned} \mathbb{E}_H[C(\boldsymbol{H}, \boldsymbol{Q}(l^*))] &= \mathbb{E}_H[\log_2(1 + \rho X_{l^*})] \\ &= \mathbb{E}_{X_{(m)}}\left[\log_2\left(1 + \rho X_{(m)}\right)\right]. \end{aligned} \tag{4}$$

We note that other measures of performance can also be derived based on the distribution of $X_{(m)}$.

### III. ORDER STATISTICS OF THE CHI-SQUARE DISTRIBUTION

For any $n \geq 1$, the probability density function (pdf) and the cumulative distribution function (cdf) of the chi-square random variable $X_l \geq 0$ with $2n$ degrees of freedom are respectively given by

$$f(x) = e^{-x} \frac{x^{n-1}}{(n-1)!} \tag{5}$$

and

$$F(x) = 1 - e^{-x} \sum_{k=0}^{n-1} \frac{x^k}{k!}, \tag{6}$$

for $x \geq 0$. Clearly, the upper endpoint $\omega(F) \triangleq \sup\{x | F(x) < 1\}$ is infinity.

Let $F_{(m)}$ and $f_{(m)}$ denote the cdf and the pdf of $X_{(m)}$. It is well known that

$$F_{(m)}(x) \triangleq F_{X_{(m)}}(x) = F^m(x) = \left(1 - e^{-x}\sum_{k=0}^{n-1}\frac{x^k}{k!}\right)^m \tag{7}$$

and

$$\begin{aligned} f_{(m)}(x) &\triangleq f_{X_{(m)}}(x) = m \cdot F^{m-1}(x) \cdot f(x) \\ &= m\left(1 - e^{-x}\sum_{k=0}^{n-1}\frac{x^k}{k!}\right)^{m-1} e^{-x}\frac{x^{n-1}}{(n-1)!}, \end{aligned} \tag{8}$$

respectively, for $x \geq 0$ [9, pp. 9–11]. Clearly $F_{(m)}(x) = 0$ and $f_{(m)}(x) = 0$ for $x < 0$. The ergodic capacity of transmit antenna selection can be computed from the above, by numerical integration of (4). Analytical solutions seem to be hard to obtain and offer no insights into channel hardening and scheduling gain for such a system.

## A. A Key Convergence Result

First, we have the following technical result whose proof may be found in the Appendix.

*Lemma 1:* Let $F(\cdot)$ and $\omega(F)$ be as defined above, then

1) For any $t > 0$, the value $R(t) \triangleq \int_t^{\omega(F)}(1 - F(y))\mathrm{d}y/\{1 - F(t)\}$ is well defined and satisfies $1 \leq R(t) < \infty$. As $t \to \infty$,

$$R(t) = \begin{cases} 1, & \text{if } n = 1, \\ 1 + (n-1)\frac{1}{t} + O\left(\frac{1}{t^2}\right), & \text{if } n \geq 2. \end{cases} \tag{9}$$

2) $F(\cdot)$ is in the domain of attraction of the *Gumbel* distribution. That is to say, for all fixed real $x$, as $m \to \infty$,

$$F^m(a_m + b_m x) \to G(x) \triangleq \exp(-e^{-x}), \tag{10}$$

where $G(x)$ is the Gumbel cdf, and the normalizing constants $a_m$ and $b_m$ can be selected to be

$$a_m = q_m \triangleq F^{-1}(1 - 1/m) \tag{11}$$

$$b_m = R(q_m). \tag{12}$$

□

In the above lemma, other choices of $a_m$ and $b_m$ are also possible. We will study this next.

*Notation:* For any two real-valued sequences $c_m$ and $d_m$, we define

$$c_m \approx d_m \quad \text{if and only if} \quad \lim_{m\to\infty}|c_m - d_m| = 0;$$

$$c_m = \Theta(d_m) \quad \text{if and only if} \quad 0 < \lim_{m\to\infty} c_m/d_m < \infty.$$

*Lemma 2:* For $F(\cdot)$ and $G(\cdot)$ defined as above,

$$\lim_{m\to\infty} F^m(a_m + b_m x) = G(x)$$

if and only if the normalizing constants $a_m$ and $b_m$ satisfy

$$a_m \approx \ln m + (n-1)\ln(\ln m) - \ln(n-1)! \tag{13}$$

$$b_m \approx 1. \tag{14}$$

□

*Theorem 3:* The variance of $X_{(m)}$ is bounded away form 0, i.e., the effective channel will exhibit significant fluctuations.

*Proof:* It is known that the Gumbel distribution has a mean $\gamma = 0.5772...$ (Euler's constant) and variance $\pi^2/6$ ([9, p. 298]). Thus the mean and variance of $X_{(m)}$ will approach to $a_m + \gamma$ and $b_m^2 \pi^2/6$, respectively, as $m$ increases. Hence, we have

$$\mathbb{E}\left[X_{(m)}\right] \approx a_m + \gamma, \tag{15}$$

$$\mathrm{Var}\left[X_{(m)}\right] \approx \frac{\pi^2}{6}. \tag{16}$$

Because the variance is neither zero for any $m$ nor goes near zero as $m$ increases, the channel will fluctuate considerably. ∎

This suggest that the scheduling gain may go to zero considerably slower than for MIMO. The scheduling gain of the above antenna-selection system is computed in Section V.

## B. Optimizing the Rate of Convergence

In this subsection, we study the best pairs of normalizing constants $a_m$ and $b_m$ that provide an accurate fit to the Gumbel distribution, even for relatively small values of $m$. For example, the simple choice of $a_m = \ln m + (n-1)\ln(\ln m) - \ln(n-1)!$ and $b_m = 1$ yields a poor estimate of the distribution for small $m$, and may not yield accurate results for a realistic numbers of transmit antennas.

It is convenient to first introduce the sequence

$$\alpha_m \triangleq \max\left\{x \geq 0 \,\bigg|\, e^{-x}\frac{x^{n-1}}{(n-1)!} = \frac{1}{m}\right\}, \tag{17}$$

which is well defined for all $m$ greater than some $M(n)$. Clearly, $\alpha_m \to \infty$ as $m \to \infty$.

*Theorem 4:* Let $F(\cdot)$ and $G(\cdot)$ be as given above. Consider the rate of convergence (for fixed $x$) of

$$|F^m(a_m + b_m x) - G(x)|, \tag{18}$$

as $m \to \infty$.

1) The optimal sequences of constants $a_m$ and $b_m$ minimizing this rate is

$$a_m = \begin{cases} \alpha_m, & \text{if } n = 1, \\ \alpha_m + (n-1)\frac{1}{\alpha_m} + O\left(\frac{1}{\alpha_m^2}\right), & \text{if } n \geq 2, \end{cases} \quad (19)$$

and

$$b_m = \begin{cases} 1, & \text{if } n = 1, \\ 1 + (n-1)\frac{1}{\alpha_m} + O\left(\frac{1}{\alpha_m^2}\right), & \text{if } n \geq 2. \end{cases} \quad (20)$$

2) For $n = 1$, the optimal rate of convergence for any given $x$ is

$$|F^m(a_m + b_m x) - G(x)| = \Theta\left(\frac{1}{m}\right), \quad (21)$$

and for $n \geq 2$,

$$|F^m(a_m + b_m x) - G(x)| = \Theta\left(\frac{1}{(\log m)^2}\right). \quad (22)$$

□

*Corollary 5:* For $F(\cdot)$, $G(\cdot)$, and $q_m$ as given above, the choice of $a_m = q_m$ and

$$b_m = \begin{cases} 1, & \text{if } n = 1, \\ 1 + (n-1)\frac{1}{q_m} + O\left(\frac{1}{q_m^2}\right), & \text{if } n \geq 2 \end{cases} \quad (23)$$

also satisfies (19) and (20), and this is therefore optimal in the sense of minimizing the rate of convergence defined in (18). □

The proofs may be found in the Appendix.

From (9) and Corollary 5, we can check that the choice of normalizing constants in Lemma 1 ($a_m = q_m$, $b_m = R(q_m)$) is optimal. Fig. 1 shows that the Gumbel approximation is an excellent fit with this choice of normalization coefficients. It can also be seen from the figure that the variance of $X_{(m)}$ stays bounded away from zero as $m \to \infty$.

## C. Outage Capacity

By using the above results, given a rate $C_0$, the corresponding outage probability $P_{\text{out}}(C_0)$ can be approximated by

$$\begin{aligned} P_{\text{out}}(C_0) &\triangleq \Pr\left\{\log_2(1 + \rho X_{(m)}) \leq C_0\right\} \\ &\approx G\left(\frac{\frac{2^{C_0}-1}{\rho} - a_m}{b_m}\right). \end{aligned} \quad (24)$$

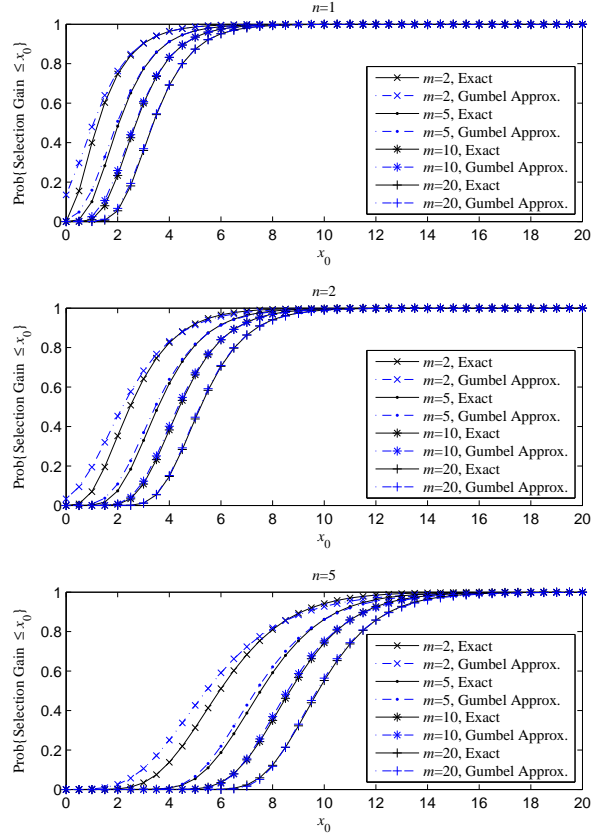

Fig. 1. Exact and Gumbel-approximate ($a_m = q_m$, $b_m = R(q_m)$) distributions of the selection gain $X_{(m)}$ for $m = 2, 5, 10, 20$ and $n = 1, 2, 5$.

The outage capacity $C_{\text{out}}(P_0)$ can then be approximated as

$$\begin{aligned} C_{\text{out}}(P_0) &\triangleq P_{\text{out}}^{-1}(P_0) = \log_2[1 + \rho F^{-1}(P_0^{\frac{1}{m}})] \\ &\approx \log_2[1 + \rho\{a_m - b_m \ln(-\ln P_0)\}]. \end{aligned} \quad (25)$$

Fig. 2 shows 10% outage capacity of transmit antenna selection and MIMO without feedback for SINR $\rho = 5$ dB. The normalizing constants for the approximations are chosen to be the same as those in Fig. 1. The figure indicates that the above approximations improve as $m$ increases. Additionally, even ignoring the scheduling gain, the *transmit antenna selection scheme outperforms full open loop MIMO schemes* in terms of outage capacity, when the number of receive antennas is small.

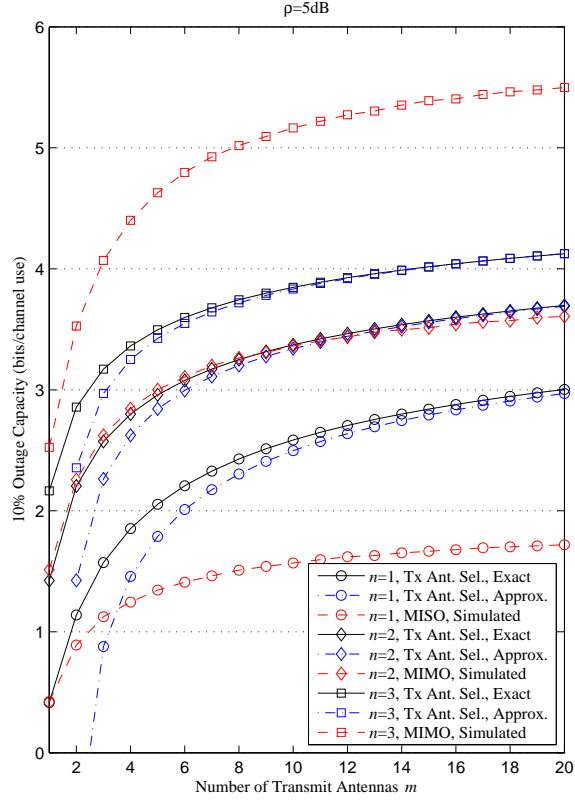

Fig. 2. 10% outage capacity of transmit antenna selection and MIMO versus $m$ for $n = 1, 2, 3$ at $\rho = 5$dB.

## IV. ERGODIC CAPACITY

### A. Some Useful Bounds

*Theorem 6:* For a random variable $X$ with cdf $F(\cdot)$ as above,

$$q_m \leq \mathbb{E}[X_{(m)}] \leq q_{e^\gamma(m+1)}, \tag{26}$$

where $\gamma$ is Euler's constant ($e^\gamma = 1.7810...$), $q_m$ is the quantile defined in (11), and $q_{e^\gamma(m+1)} \triangleq F^{-1}\left(1 - 1/e^\gamma(m+1)\right)$. The ergodic capacity $\bar{C}(\rho) \triangleq \mathbb{E}\left[\log_2\left(1 + \rho\, X_{(m)}\right)\right]$ also satisfies

$$\log_2\left(1 + \rho\, q_m\right) \leq \bar{C}(\rho) \leq \log_2\left(1 + \rho\, q_{e^\gamma(m+1)}\right), \tag{27}$$

for any $\rho > 0$. □

The proof may be found in the Appendix.

From Jensen's inequality and the left inequality in (26), we obtain

$$\log_2\left(1 + \rho\, q_m\right) \leq \bar{C}(\rho) \leq \log_2\left(1 + \rho\, \mathbb{E}[X_{(m)}]\right) \leq \log_2\left(1 + \rho\, q_{e^\gamma(m+1)}\right) \tag{28}$$

From the fact that $q_{e^\gamma(m+1)} - q_m \approx \gamma$, we can see that

$$\log_2\left(1 + \rho\, q_{e^\gamma(m+1)}\right) \approx \log_2\left(1 + \rho\, q_m\right). \tag{29}$$

It follows that

$$\bar{C}(\rho) \approx \log_2\left(1 + \rho\, \mathbb{E}[X_{(m)}]\right) \tag{30}$$

$$\approx \log_2\left(1 + \rho\, (q_m + \gamma)\right). \tag{31}$$

### B. Asymptotic Analysis for Large Number of Receive Antennas

A chi-square random variable $X$ with cdf $F(\cdot)$ as in above is the sum of $2n$ i.i.d. random variables with mean and variance $= 0.5$. In order to study the change of $F(\cdot)$ and $q_m$ as functions of $n$, it will be convenient to write $F_n(\cdot)$ and $q_m(n)$ respectively. As $n$ increases, (from the central limit theorem) $(X - n)/\sqrt{n}$ converges to the Gaussian distribution with mean zero and variance one. Thus, for large $n$,

$$q_m(n) = F_n^{-1}\left(1 - \frac{1}{m}\right) = n + O(\sqrt{n}). \tag{32}$$

By Theorem 6 and (32),

$$\mathbb{E}[X_{(m)}] = n + O(\sqrt{n}), \tag{33}$$

$$\bar{C}(\rho) = \log_2\left(1 + \rho\, (n + O(\sqrt{n}))\right). \tag{34}$$

### C. Numerical Results

Fig. 3 demonstrates the ergodic capacity of various systems for a few basic antenna configurations. It is observed that the transmit antenna selection scheme performance is superior to an open loop MIMO scheme in the low SINR regime. *This may have some implications on wireless system design* as most of the users in modern wireless systems have low SINRs. In fact, the ergodic capacity of open loop MIMO is upper bounded by $n \log_2(1 + \rho)$, while the ergodic capacity of transmit antenna selection goes to infinity as $m \to \infty$, and is not upper bounded.

In Figures 4 and 5, we study the ergodic capacity respectively as a function $m$ and $n$, for SINR $\rho = 5$ dB. It is observed that our bounds in (27) and approximation in (31) for the ergodic capacity of transmit antenna selection schemes are very good and become exact as $m$ increases. In terms of the ergodic capacity, it is also seen that transmit antenna selection outperforms open loop MIMO when the number of receive antennas is small. If more than two receive antennas are deployed, it appears that open loop MIMO is better than transmit antenna selection for a moderate number of transmit antennas ($m = 1, \cdots, 20$) at SINR $\rho = 5$ dB. Nonetheless, in lower SINR regimes, transmit antenna selection will outperform open loop MIMO even for more than two receive antennas.

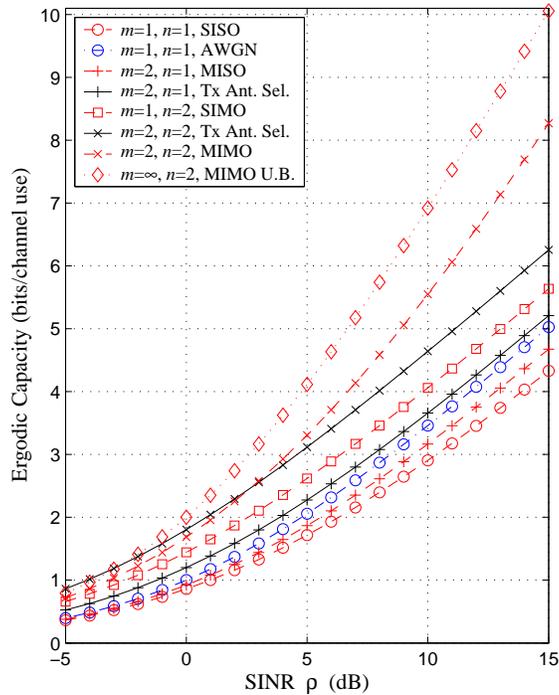

Fig. 3. Ergodic capacity versus $\rho$ for $m = 1, 2$ and $n = 1, 2$.

## V. Scheduling

For a single cell with multiple users, many scheduling strategies have been proposed. Among them, it is known that a greedy scheduling algorithm maximizes the total system capacity. In greedy scheduling, the base station selects the user with the best channel at any given time. Only this user may communicate with the base station.

It is known that multiple transmit antennas in MIMO reduce channel fluctuations and thus the benefits of scheduling decrease as the number of transmit antennas increases [8]. However, that is not necessarily the case for transmit antenna selection because there are significant channel fluctuations even after deploying a large number of transmit antennas. We compare the system capacity of greedy scheduling for antenna selection to that of round robin scheduling for antenna selection as well as greedy and round-robin scheduling for MIMO. Our basic assumption for the analysis is that all users have the same number of antennas.

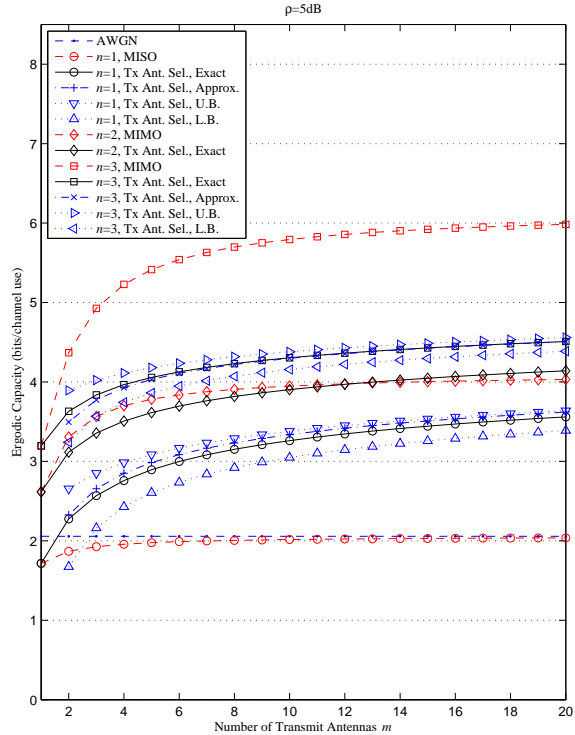

Fig. 4. Ergodic capacity versus $m$ for $n = 1, 2, 3$ at $\rho = 5$ dB.

We define the *capacity of a scheduling algorithm* where there are $K$ users and each user is equipped with $n$ antennas in case of downlink or $m$ antennas in case of uplink as the average system capacity after scheduling.

The greedy scheduling capacity is then the same as the ergodic capacity of transmit-antenna-selection with $mK$ transmit antennas, and it is given by

$$\mathbb{E}\left[\log_2\left(1 + \rho\, X_{(mK)}\right)\right] \approx \log_2(1 + \rho\,(q_{mK} + \gamma)), \tag{35}$$

using (31). From (27), this is upper and lower bounded by $\log_2(1 + \rho\, q_{mK})$ and $\log_2(1 + \rho\, q_{e^\gamma(mK+1)})$, respectively. Note that round robin scheduling has the same capacity as the ergodic capacity of a point-to-point link with the same number of transmit antennas and is given by

$$\mathbb{E}\left[\log_2\left(1 + \rho\, X_{(m)}\right)\right] \approx \log_2(1 + \rho\,(q_m + \gamma)), \tag{36}$$

and this is upper and lower bounded in (27).

Fig. 6 shows the average system capacity of transmit-antenna-selection and MIMO versus $m$. The approximations and bounds in this figure are those given in (35), (36), and the discussions following them. We can check that the capacity of greedy scheduling for transmit-antenna-selection increases as the number of transmit antennas increases while it decreases for MIMO. Thus, greedy scheduling works well with transmit antenna selection in the presence of a large number of transmit antennas.

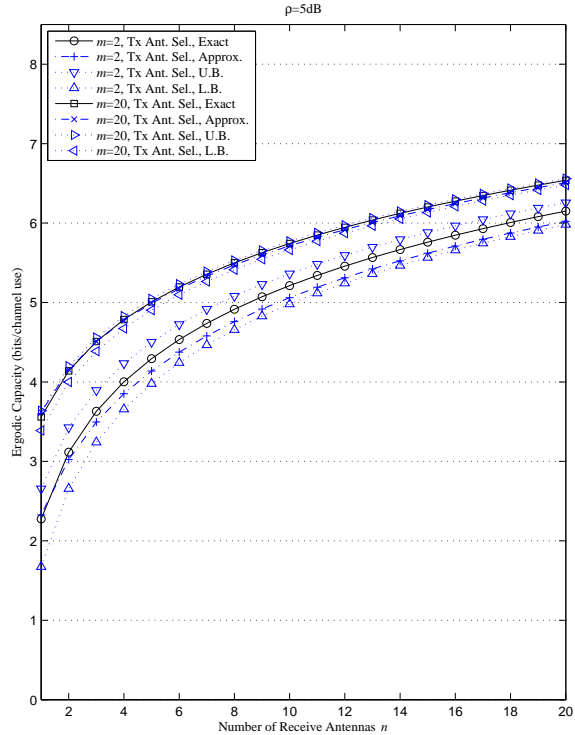

Fig. 5. Ergodic capacity versus $n$ for $m = 2, 20$ at $\rho = 5$ dB.

We define the *scheduling gain* as the increase of average system capacity over that without scheduling. It can be approximated as

$$\mathbb{E}\left[\log_2\left(1 + \rho\, X_{(mK)}\right)\right] - \mathbb{E}\left[\log_2\left(1 + \rho\, X_{(m)}\right)\right] \tag{37}$$

$$\approx \log_2\left(\frac{1 + \rho\,(q_{mK} + \gamma)}{1 + \rho\,(q_m + \gamma)}\right) = \log_2\left(1 + \rho\,\frac{q_{mK} - q_m}{1 + \rho\,(q_m + \gamma)}\right) \tag{38}$$

$$\approx \log_2\left(1 + \frac{q_{mK} - q_m}{q_m}\right) \approx \log_2\left(1 + \frac{\ln K}{q_m}\right) \tag{39}$$

$$= O\left(\frac{1}{q_m}\right) = O\left(\frac{1}{\log m}\right). \tag{40}$$

The numerically integrated values of (37) for $1 \leq m \leq 20$ and the approximated values of (38) for $2 \leq m \leq 20$ are tabulated in Table I for $-5\text{dB} \leq \rho \leq 10\text{dB}$ with 5dB increament, $n = 1$, and $K = 32$. Note that the approximations are at most 0.1 bits away from the exact values. Thus, (38) is a good

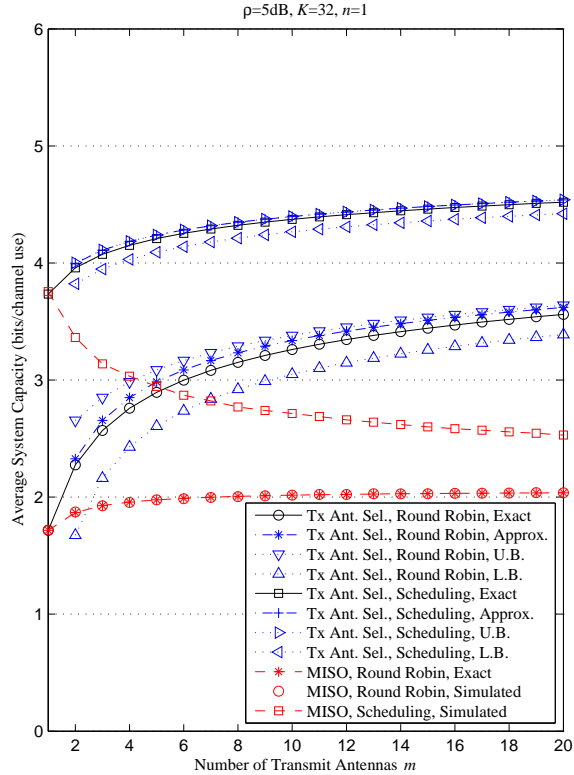

Fig. 6. Average system capacity after scheduling versus $m$ for $n = 1$ and $K = 32$ at $\rho = 5$dB.

approximation to (37).

We define the *fractional scheduling gain* of greedy scheduling to be the greedy scheduling gain of as a fraction of the capacity of round robin scheduling. Since $\mathbb{E}\left[\log_2\left(1 + \rho\, X_{(m)}\right)\right] \approx \log_2(1 + \rho\,(q_m + \gamma))$, it is easy to see that

$$\frac{\mathbb{E}\left[\log_2\left(1 + \rho\, X_{(mK)}\right)\right] - \mathbb{E}\left[\log_2\left(1 + \rho\, X_{(m)}\right)\right]}{\mathbb{E}\left[\log_2\left(1 + \rho\, X_{(m)}\right)\right]} = O\left(\frac{1}{\log m \, \log(\log m)}\right). \tag{41}$$

## VI. CONCLUSION

In this work, we considered the use of transmit antenna selection in multiple antenna wireless systems. It was shown that for antenna selection systems (unlike MIMO systems), the channel hardens at much slower rate, and thus significant multiuser scheduling gain can exist. Additionally, it was shown that even without this scheduling gain, transmit antenna selection systems outperform open loop MIMO systems at low SINR regimes, particularly for a small number of receive antennas. The implications of these results on wireless system design was briefly discussed.

$K = 32, n = 1$

| m \ ρ | -5dB | 0dB | 5dB | 10dB |
|---|---|---|---|---|
| 1 | 0.8084 | 1.4366 | 2.0183 | 2.4095 |
| 2 | 0.7803 (0.83) | 1.2899 (1.34) | 1.6826 (1.67) | 1.8935 (1.82) |
| 3 | 0.7528 (0.78) | 1.1958 (1.20) | 1.5048 (1.45) | 1.6540 (1.56) |
| 4 | 0.7308 (0.75) | 1.1308 (1.12) | 1.3926 (1.33) | 1.5119 (1.42) |
| 5 | 0.7131 (0.72) | 1.0826 (1.06) | 1.3139 (1.25) | 1.4155 (1.33) |
| 6 | 0.6984 (0.70) | 1.0449 (1.02) | 1.2548 (1.20) | 1.3445 (1.27) |
| 7 | 0.6861 (0.69) | 1.0144 (0.99) | 1.2082 (1.15) | 1.2895 (1.21) |
| 8 | 0.6754 (0.67) | 0.9890 (0.96) | 1.1702 (1.11) | 1.2464 (1.17) |
| 9 | 0.6661 (0.66) | 0.9673 (0.94) | 1.1384 (1.09) | 1.2084 (1.14) |
| 10 | 0.6578 (0.65) | 0.9485 (0.92) | 1.1113 (1.06) | 1.1772 (1.11) |
| 11 | 0.6504 (0.65) | 0.9321 (0.90) | 1.0877 (1.04) | 1.1502 (1.09) |
| 12 | 0.6438 (0.64) | 0.9174 (0.89) | 1.0670 (1.02) | 1.1267 (1.07) |
| 13 | 0.6377 (0.63) | 0.9045 (0.88) | 1.0486 (1.00) | 1.1058 (1.05) |
| 14 | 0.6321 (0.63) | 0.8925 (0.87) | 1.0321 (0.99) | 1.0872 (1.03) |
| 15 | 0.6270 (0.62) | 0.8816 (0.86) | 1.0172 (0.97) | 1.0704 (1.02) |
| 16 | 0.6222 (0.62) | 0.8717 (0.85) | 1.0035 (0.96) | 1.0551 (1.00) |
| 17 | 0.6178 (0.61) | 0.8626 (0.84) | 0.9911 (0.95) | 1.0411 (0.99) |
| 18 | 0.6137 (0.61) | 0.8541 (0.83) | 0.9796 (0.94) | 1.0283 (0.98) |
| 19 | 0.6098 (0.60) | 0.8463 (0.82) | 0.9690 (0.93) | 1.0164 (0.97) |
| 20 | 0.6062 (0.60) | 0.8390 (0.81) | 0.9591 (0.92) | 1.0054 (0.96) |

TABLE I. Exact and approximated (parenthesis) scheduling gain in bits for various $m$ and $\rho$ for $n = 1$ and $K = 32$.

## Appendix

*Proof of Lemma 1:* In the first part of the lemma, for $t > 0$,

$$\int_t^\infty (1 - F(y))\mathrm{d}y = \sum_{k=0}^{n-1} \int_t^\infty e^{-y} \frac{y^k}{k!} \mathrm{d}y$$

$$= e^{-t} \sum_{k=0}^{n-1} \sum_{i=0}^{k} \frac{t^i}{i!} < \infty \quad (42)$$

and thus

$$1 \leq R(t) \triangleq \frac{\int_t^\infty (1 - F(y))\mathrm{d}y}{1 - F(t)} = \frac{e^{-t} \sum_{k=0}^{n-1} \sum_{i=0}^{k} \frac{t^i}{i!}}{e^{-t} \sum_{k=0}^{n-1} \frac{t^k}{k!}} < \infty. \quad (43)$$

From (43), if $n = 1$, $R(t) = 1$, and if $n \geq 2$,

$$R(t) = \frac{\frac{t^{n-1}}{(n-1)!} + 2\frac{t^{n-2}}{(n-2)!} + O(t^{n-3})}{\frac{t^{n-1}}{(n-1)!} + \frac{t^{n-2}}{(n-2)!} + O(t^{n-3})}$$

$$= 1 + (n-1)\frac{1}{t} + O\left(\frac{1}{t^2}\right). \tag{44}$$

This proves the first part of the lemma.

From (44), $\lim_{t\to\infty} xR(t)/t = 0$ for all real $x$ and thus

$$\lim_{t\to\infty} \frac{1 - F(t + xR(t))}{1 - F(t)} = \lim_{t\to\infty} \frac{e^{-(t+xR(t))}(t + xR(t))^{(n-1)}}{e^{-t}t^{n-1}}$$

$$= \lim_{t\to\infty} e^{-xR(t)}\left(1 + \frac{xR(t)}{t}\right)^{n-1}$$

$$= e^{-x}. \tag{45}$$

For the second part, the result follows from (45) and the following theorem whose proof can be found in [10, Ch. 2]. ∎

*Theorem 7:* Define the upper endpoint $\omega(F) \triangleq \sup\{x|F(x) < 1\}$. Then $F$ is in the domain of attraction of $G = \exp(-\exp(-x))$ if and only if there exists some finite $a < \omega(F)$ such that

$$\int_a^{\omega(F)} (1 - F(y))\mathrm{d}y < \infty, \tag{46}$$

and for all real $x$,

$$\lim_{t\to\omega(F)} \frac{1 - F(t + xR(t))}{1 - F(t)} = e^{-x}, \tag{47}$$

where

$$R(t) \triangleq \frac{\int_t^{\omega(F)}(1 - F(y))\mathrm{d}y}{1 - F(t)}. \tag{48}$$

Moreover, the normalizing constants $a_m$ and $b_m$ in (10) can be chosen as

$$a_m = q_m^* \triangleq \inf\left\{x \,\Big|\, 1 - F(x) \leq \frac{1}{m}\right\} \tag{49}$$

and

$$b_m = R(q_m^*). \tag{50}$$

□

The following lemma in [10, Ch. 2] will also prove useful in the proof of Lemma 2.

*Lemma 8:* Let $X_m$ be any sequence of random variables such that, for some constants $a_m$, $a_m^*$, $b_m > 0$, $b_m^* > 0$,

$$\lim_{m\to\infty} \Pr\{X_m \leq a_m + b_m x\} = \lim_{m\to\infty} \Pr\{X_m \leq a_m^* + b_m^* x\} = G(x), \tag{51}$$

for all continuity points of the distribution function $G(x)$. Then (51) holds if and only if both of the following hold

$$\lim_{m \to \infty} \frac{a_m - a_m^*}{b_m} = 0 \tag{52}$$

$$\lim_{m \to \infty} \frac{b_m^*}{b_m} = 1. \tag{53}$$

□

*Proof of Lemma 2:* We prove (14) first. From Lemma 8 necessary and sufficient conditions for $b_m$ are

$$\lim_{m \to \infty} \frac{b_m}{R(q_m)} = 1 \tag{54}$$

From (9) in Lemma 1 we know that

$$R(t) = \begin{cases} 1, & \text{if } n = 1, \\ 1 + (n-1)\frac{1}{t} + O\left(\frac{1}{t^2}\right), & \text{if } n \geq 2, \end{cases} \tag{55}$$

and

$$\lim_{m \to \infty} q_m = \infty. \tag{56}$$

Thus $\lim_{m \to \infty} b_m$ should be 1. Since the converse is also true, we have proved (14). Because of (14), (52) in Lemma 8 reduces to

$$\lim_{m \to \infty} (a_m - a_m^*) = 0 \tag{57}$$

and the conditions for $a_m$ and $b_m$ can be separated. Since the reference $a_m$ can be chosen to be $q_m$ in this case, it suffices to prove that

$$\lim_{m \to \infty} [q_m - \{\ln m + (n-1)\ln(\ln m) - \ln(n-1)!\}] = 0. \tag{58}$$

From the definition of $q_m$,

$$e^{-q_m} \sum_{k=0}^{n-1} \frac{q_m^k}{k!} = \frac{1}{m}. \tag{59}$$

For $n = 1$,

$$q_m = \ln m, \tag{60}$$

and obviously (60) satisfies (58). Now for $n \geq 2$ and $m$ large enough that $q_m \geq 1$,

$$\sum_{k=0}^{n-1} \frac{q_m^k}{k!} = \frac{q_m^{n-1}}{(n-1)!} + \sum_{k=0}^{n-2} \frac{q_m^k}{k!}$$

$$\leq \frac{q_m^{n-1}}{(n-1)!} + q_m^{n-2} \sum_{k=0}^{n-2} \frac{1}{k!}$$

$$< \frac{q_m^{n-1}}{(n-1)!} + q_m^{n-2} e. \tag{61}$$

Thus,
$$e^{-q_m} \frac{q_m^{n-1}}{(n-1)!} \leq \frac{1}{m} < e^{-q_m} \left\{ \frac{q_m^{n-1}}{(n-1)!} + e \cdot q_m^{n-2} \right\}. \tag{62}$$

Taking the logarithms on both sides of (62) gives
$$-q_m + (n-1)\ln q_m - \ln(n-1)! \leq -\ln m$$
$$< -q_m + \ln\left[\frac{q_m^{n-1}}{(n-1)!}\left(1 + \frac{e(n-1)!}{q_m}\right)\right]$$
$$= -q_m + (n-1)\ln q_m - \ln(n-1)! + \ln\left(1 + \frac{e(n-1)!}{q_m}\right) \tag{63}$$

and it yields
$$\ln m \leq q_m - (n-1)\ln q_m + \ln(n-1)! < \ln m + \ln\left(1 + \frac{e(n-1)!}{q_m}\right). \tag{64}$$

Define $\varepsilon_m$ as
$$q_m = \ln m + (n-1)\ln(\ln m) - \ln(n-1)! + \varepsilon_m \tag{65}$$

and then we must prove $\lim_{m \to \infty} \varepsilon_m = 0$. From (64),
$$\lim_{m \to \infty} \frac{q_m}{\ln m} = 1 \quad \text{and} \quad \lim_{m \to \infty} \frac{\ln q_m}{\ln m} = 0. \tag{66}$$

We can see
$$\lim_{m \to \infty} \frac{\varepsilon_m}{\ln m} = 0. \tag{67}$$

Again, by substituting (65) for the leftmost $q_m$ in (64) and defining $\delta_m \triangleq \ln q_m - \ln(\ln m)$,
$$(n-1)\delta_m \leq \varepsilon_m < (n-1)\delta_m + \ln\left(1 + \frac{e(n-1)!}{q_m}\right).$$

Because
$$\lim_{m \to \infty} \ln\left(1 + \frac{e(n-1)!}{q_m}\right) = 0, \tag{68}$$

we only need to show $\lim_{m \to \infty} \delta_m = 0$. For $m$ such that $q_m \neq \ln m$,
$$\delta_m = \ln q_m - \ln(\ln m)$$
$$= (q_m - \ln m)\frac{1}{\zeta_m}, \quad \text{for some } \zeta_m \in (q_m, \ln m) \text{ or } (\ln m, q_m)$$
$$= \frac{q_m - \ln m}{\ln m + \eta_m(q_m - \ln m)}, \quad \text{for some } \eta_m \in (0, 1)$$
$$= \frac{(n-1)\ln(\ln m) - \ln(n-1)! + \varepsilon_m}{\ln m + \eta_m\{(n-1)\ln(\ln m) - \ln(n-1)! + \varepsilon_m\}}, \tag{69}$$

by mean value theorem. From (67) and (69),
$$\lim_{m \to \infty} \delta_m = 0. \tag{70}$$

■

*Proof of Theorem 4:* First, note that we borrow some of the proof techniques from Hall's paper [12] and Galambos's book [10, Sec. 2.10]. We see that $\alpha_m$ satisfies

$$\alpha_m - (n-1)\ln \alpha_m + \ln(n-1)! = \ln m. \tag{71}$$

As in the proof of Lemma 2, it can be shown that

$$\lim_{m \to \infty} [\alpha_m - \{\ln m + (n-1)\ln(\ln m) - \ln(n-1)!\}] = 0. \tag{72}$$

Then, we can express the general normalizing constants $a_m$ and $b_m$ as

$$a_m = \alpha_m + \delta_m \quad \text{and} \quad b_m = 1 + \varepsilon_m, \tag{73}$$

where $\delta_m$ and $\varepsilon_m$ are sequences satisfy

$$\lim_{m \to \infty} \delta_m = 0 \quad \text{and} \quad \lim_{m \to \infty} \varepsilon_m = 0. \tag{74}$$

For $n = 1$, (21) and the choice of (19) and (20) are proved in [10, p. 142] because the chi-square distribution just become the exponential distribution. Hence, let us assume $n \geq 2$, and define

$$z_m(x) \triangleq m[1 - F(a_m + b_m x)]. \tag{75}$$

We will shortly prove that $z_m(x) \to e^{-x}$ as $m \to \infty$ but assuming that this is the case, from (75),

$$F^m(a_m + b_m x) = \left[1 - \frac{z_m(x)}{m}\right]^m, \tag{76}$$

and by the triangle inequality,

$$\left||F^m(a_m + b_m x) - e^{-z_m(x)}| - |e^{-z_m(x)} - G(x)|\right| \leq |F^m(a_m + b_m x) - G(x)|$$
$$\leq |F^m(a_m + b_m x) - e^{-z_m(x)}| + |e^{-z_m(x)} - G(x)|. \tag{77}$$

Also from [10, p. 8], for any $z \in (0, 1/2)$,

$$e^{-mz} - (1-z)^m[e^{2mz^2} - 1] < (1-z)^m \leq e^{-mz}. \tag{78}$$

For fixed $x$, because $z_m(x) \to e^{-x}$, $z_m(x)/m \in (0, 1/2)$ for large enough $m$. By (78) with $z = z_m(x)/m$,

$$\begin{aligned}
|F^m(a_m + b_m x) - e^{-z_m(x)}| &\leq \left[1 - \frac{z_m(x)}{m}\right]^m \left[\exp\left(\frac{2z_m^2(x)}{m}\right) - 1\right] \\
&\leq e^{-z_m(x)} \left[\exp\left(\frac{2z_m^2(x)}{m}\right) - 1\right] \\
&= e^{-z_m(x)} \left[\frac{2z_m^2(x)}{m} + O\left(\frac{1}{m^2}\right)\right]. \tag{79}
\end{aligned}$$

The rate of convergence of $|F^m(a_m+b_m x)-e^{-z_m(x)}|$ is dominated by the $1/m$ term, which will turn out to be much faster than that of $|e^{-z_m(x)}-G(x)|$. Now, consider the rate of convergence of $|e^{-z_m(x)}-G(x)|$. From the definition of $F(x)$ in (6), we can easily see that as $x \to \infty$,

$$1 - F(x) = e^{-x} \frac{x^{n-1}}{(n-1)!}\left[1 + \frac{n-1}{x} + O\left(\frac{1}{x^2}\right)\right]. \tag{80}$$

From (17) and noting that $\delta_m, \varepsilon_m, 1/\alpha_m \to 0$ as $m \to \infty$,

$$e^{-(a_m+b_m x)}\frac{(a_m+b_m x)^{n-1}}{(n-1)!}$$

$$= e^{-\alpha_m} \frac{\alpha_m^{n-1}}{(n-1)!} e^{-x} e^{-(\delta_m+\varepsilon_m x)}\left(1 + \frac{x}{\alpha_m} + \frac{\delta_m}{\alpha_m} + \frac{\varepsilon_m}{\alpha_m}x\right)^{n-1}$$

$$= \frac{1}{m}e^{-x}\left[1 - (\delta_m+\varepsilon_m x) + O(\delta_m^2+\varepsilon_m^2)\right]$$

$$\cdot \left[1 + (n-1)\frac{x}{\alpha_m} + O\left(\frac{1}{\alpha_m^2} + \frac{\delta_m}{\alpha_m} + \frac{\varepsilon_m}{\alpha_m}\right)\right]$$

$$= \frac{1}{m}e^{-x}\left[1 - (\delta_m+\varepsilon_m x) + \frac{n-1}{\alpha_m}x + O\left(\frac{1}{\alpha_m^2} + \delta_m^2 + \varepsilon_m^2\right)\right] \tag{81}$$

and

$$1 + \frac{n-1}{a_m+b_m x} + O\left(\frac{1}{(a_m+b_m x)^2}\right) = 1 + \frac{n-1}{\alpha_m} + O\left(\frac{1}{\alpha_m^2}\right). \tag{82}$$

Because $1 - F(a_m+b_m x)$ is equal to the product of (81) and (82),

$$1 - F(a_m+b_m x)$$
$$= \frac{1}{m}e^{-x}\left[1 - (\delta_m+\varepsilon_m x) + \frac{(n-1)}{\alpha_m}(1+x) + O\left(\frac{1}{\alpha_m^2} + \delta_m^2 + \varepsilon_m^2\right)\right]. \tag{83}$$

From this, it is clear that as $m \to \infty$,

$$z_m(x) = m[1 - F(a_m+b_m x)] \to e^{-x}. \tag{84}$$

Using the fact that $G(x) = \exp(-e^{-x})$,

$$|e^{-z_m(x)} - G(x)|$$
$$= G(x)|\exp(e^{-x} - z_m(x)) - 1|$$
$$= G(x)\left|(e^{-x} - z_m(x)) + O\left((e^{-x}-z_m(x))^2\right)\right|$$
$$= G(x)e^{-x}\left|(\delta_m+\varepsilon_m x) - \frac{(n-1)}{\alpha_m}(1+x) + O\left(\frac{1}{\alpha_m^2} + \delta_m^2 + \varepsilon_m^2\right)\right| \tag{85}$$

Obviously, to cancel out $(n-1)(x+1)/\alpha_m$,

$$\delta_m = (n-1)\frac{1}{\alpha_m} + O\left(\frac{1}{\alpha_m^2}\right) \tag{86}$$

and
$$\varepsilon_m = (n-1)\frac{1}{\alpha_m} + O\left(\frac{1}{\alpha_m^2}\right). \tag{87}$$

Therefore, (19) and (20) must be satisfied to optimize the rate of convergence. Moreover, if $\delta_m$ and $\varepsilon_m$ are chosen as (86) and (87), by the second order expansion, it can be shown that the terms of $O(1/\alpha_m^2)$ cannot be canceled out. Thus
$$|e^{-z_m(x)} - G(x)| = \Theta\left(\frac{1}{\alpha_m^2}\right). \tag{88}$$

Because
$$\alpha_m = \ln m + (n-1)\ln(\ln m) - \ln(n-1)! + o(1), \tag{89}$$

$1/\alpha_m^2 \to 0$ is much slower than $1/m \to 0$ as $m \to \infty$. Hence, combining (88) and (79) into (77) yields
$$|F^m(a_m + b_m x) - G(x)| = \Theta\left(\frac{1}{\alpha_m^2}\right). \tag{90}$$

From (89),
$$|F^m(a_m + b_m x) - G(x)| = \Theta\left(\frac{1}{(\log m)^2}\right). \tag{91}$$

∎

*Proof of Corollary 5:* For $n=1$, $q_m = \alpha_m = \ln m$ and $b_m = 1$. Therefore, (19) and (20) hold. Let us now take $n \geq 2$. From the definition of $\alpha_m$ and $q_m$,
$$\frac{1}{m} = e^{-\alpha_m}\frac{\alpha_m^{n-1}}{(n-1)!} = e^{-q_m}\frac{q_m^{n-1}}{(n-1)!}\left(1 + \sum_{k=1}^{n-1} \frac{{}_{n-1}P_k}{q_m^k}\right), \tag{92}$$

where ${}_{n-1}P_k = (n-1)!/(n-1-k)!$. By taking logarithms,
$$-\alpha_m + (n-1)\ln\alpha_m - \ln(n-1)!$$
$$= -q_m + (n-1)\ln q_m - \ln(n-1)! + \ln\left(1 + \sum_{k=1}^{n-1} \frac{{}_{n-1}P_k}{q_m^k}\right). \tag{93}$$

Define $\varepsilon_m$ as $q_m = \alpha_m + \varepsilon_m$. We can see that $\alpha_m \to \infty$ and $\varepsilon_m \to 0$ as $m \to \infty$ because of (72). Substituting $\alpha_m + \varepsilon_m$ for $q_m$ yields
$$\varepsilon_m = (n-1)\ln\left(1 + \frac{\varepsilon_m}{\alpha_m}\right) + \ln\left(1 + \sum_{k=1}^{n-1} \frac{{}_{n-1}P_k}{(\alpha_m + \varepsilon_m)^k}\right). \tag{94}$$

Define
$$\beta_m \triangleq \sum_{k=1}^{n-1} \frac{{}_{n-1}P_k}{(\alpha_m + \varepsilon_m)^k}. \tag{95}$$

It is obvious that $\varepsilon_m/\alpha_m \to 0$ and $\beta_m \to 0$ as $m \to \infty$. Thus for large enough $m$,
$$\varepsilon_m = (n-1)\left[\frac{\varepsilon_m}{\alpha_m} + O\left(\left(\frac{\varepsilon_m}{\alpha_m}\right)^2\right)\right] + \beta_m + O(\beta_m^2). \tag{96}$$

However, since

$$\beta_m = \sum_{k=1}^{n-1} \frac{{}_{n-1}P_k}{(\alpha_m + \varepsilon_m)^k} = \frac{n-1}{\alpha_m} + O\left(\frac{1}{\alpha_m^2}\right), \tag{97}$$

(96) becomes

$$\varepsilon_m = (n-1)\frac{\varepsilon_m}{\alpha_m} + \frac{n-1}{\alpha_m} + O\left(\frac{1}{\alpha_m^2}\right). \tag{98}$$

Then

$$\begin{aligned}
\varepsilon_m &= \frac{\frac{n-1}{\alpha_m} + O\left(\frac{1}{\alpha_m^2}\right)}{1 - \frac{n-1}{\alpha_m}} \\
&= \left[\frac{n-1}{\alpha_m} + O\left(\frac{1}{\alpha_m^2}\right)\right]\left[1 + \frac{n-1}{\alpha_m} + O\left(\frac{1}{\alpha_m^2}\right)\right] \\
&= (n-1)\frac{1}{\alpha_m} + O\left(\frac{1}{\alpha_m^2}\right). \tag{99}
\end{aligned}$$

Clearly, $q_m$ satisfies (19). We can obtain (20) by substituting (19) for $q_m$ in (23). ∎

*Proof of Theorem 6:* We first introduce a simple convex ordering result by van Zwet [11, Ch. 2]. Assume $X$ and $Y$ be arbitrary random variables, whose cdfs are $F_X$ and $F_Y$ respectively. Van Zwet showed that if $F_Y^{-1}(F_X(x))$ is convex, then $F_X(\mathbb{E}[X]) \leq F_Y(\mathbb{E}[Y])$ and $F_X(\mathbb{E}[X_{(m)}]) \leq F_Y(\mathbb{E}[Y_{(m)}])$, provided the expectations exist. If $F_Y^{-1}(F_X(x))$ is concave, then the inequalities are reversed. Now, we return to the chi-square distribution, where the random variable $X$ follows the cdf $F(x)$ in (6) and consider a random variable $Y$ with the cdf $F_Y(y) = -1/y$ ($-\infty < y < -1$). We can easily see that

$$F_Y(\mathbb{E}[Y_{(m)}]) = 1 - \frac{1}{m}. \tag{100}$$

Because $F_Y^{-1}(F(x)) = -1/F(x)$, we only need to show that $1/F(x)$ is convex. If so, $-1/F(x)$ is then concave and

$$\mathbb{E}[X_{(m)}] \geq F^{-1}\left(1 - \frac{1}{m}\right) = q_m. \tag{101}$$

Then, the lower bound of (26) will be proved. Because $F(0) = 0$, we can assume $X > 0$. It will be sufficient to show that

$$\frac{d^2}{dx^2}\left[\frac{1}{F(x)}\right] = -\frac{f'(x)F(x) - 2\{f(x)\}^2}{\{F(x)\}^3} > 0 \tag{102}$$

for $x > 0$. However, this can be verified explicitly when $n = 1$. For $n \geq 2$, it follows from

$$\begin{aligned}
\frac{f'(x)F(x)}{\{f(x)\}^2} &= \frac{n-1}{n}\left(1 - \frac{n!}{n-1}\sum_{k=1}^{\infty}\frac{k+1}{(k+n)!}x^k\right) \\
&< \frac{n-1}{n}. \tag{103}
\end{aligned}$$

We now show the upper bound of (26). If a random variable $Y$ has a cdf $F_Y(y) = 1 - e^{-y}$ $(0 < y < \infty)$, then

$$\mathbb{E}[Y_{(m)}] = \sum_{k=1}^{m} \frac{1}{k}. \tag{104}$$

For $x > 0$, define

$$h(x) \triangleq F_Y^{-1}(F(x)) = -\ln[1 - F(x)]. \tag{105}$$

Then, for $n = 1$, $h''(x) = 0$, and for $n \geq 2$,

$$\begin{aligned} h''(x) &= \frac{d}{dx}\left[\frac{f(x)}{1 - F(x)}\right] \\ &= \frac{\frac{x^{n-2}}{(n-2)!}\left(1 + \sum_{k=1}^{n-1}(n-1-k)\frac{x^k}{k!}\right)}{\left(\sum_{k=0}^{n-1}\frac{x^k}{k!}\right)^2} \\ &> 0. \end{aligned} \tag{106}$$

Therefore, by convex ordering,

$$\begin{aligned} F(\mathbb{E}[X_{(m)}]) &\leq F_Y(\mathbb{E}[Y_{(m)}]) = F_Y\left(\sum_{k=1}^{m}\frac{1}{k}\right) \\ &= 1 - \exp\left(-\sum_{k=1}^{m}\frac{1}{k}\right) \\ &= 1 - \frac{1}{m+1}\exp\left[-\left\{\sum_{k=1}^{m}\frac{1}{k} - \ln(m+1)\right\}\right]. \end{aligned} \tag{107}$$

For $m \geq 1$, we can easily show that $\sum_{k=1}^{m}\frac{1}{k} - \ln(m+1)$ is increasing as a function of $m$ and by the definition of $\gamma$ as Euler's constant,

$$\lim_{m \to \infty}\left[\sum_{k=1}^{m}\frac{1}{k} - \ln(m+1)\right] = \lim_{m \to \infty}\left[\left\{\sum_{k=1}^{m+1}\frac{1}{k} - \ln(m+1)\right\} - \frac{1}{m+1}\right] = \gamma. \tag{108}$$

Thus $\sum_{k=1}^{m}\frac{1}{k} - \ln(m+1) \leq \gamma$ and it yields

$$\mathbb{E}[X_{(m)}] \leq F^{-1}\left(1 - \frac{1}{e^\gamma(m+1)}\right) = q_{e^\gamma(m+1)}. \tag{109}$$

Hence, (26) is proved. The upper bound of (27) can be deduced from the upper bound of (26) by Jensen's inequality because $\rho > 0$ and $\log_2(1 + \rho(\cdot))$ is then a concave function. Now, only the proof for the lower bound of (27) remains. Define $Z \triangleq \log_2(1 + \rho X)$. Then $Z_{(m)} = \log_2(1 + \rho X_{(m)})$. The cdf of $Z$ is

$$F_Z(z) = F\left(\frac{2^z - 1}{\rho}\right). \tag{110}$$

We will show that $1/F_Z(z)$ is convex. Let $1/F(x)$ be $Q(x)$ and then

$$\frac{d^2}{dz^2}\left[\frac{1}{F_Z(z)}\right] = \left[Q''\left(\frac{2^z-1}{\rho}\right)\frac{2^z}{\rho} + Q'\left(\frac{2^z-1}{\rho}\right)\right]\frac{2^z}{\rho}(\ln 2)^2. \quad (111)$$

If we substitute $(2^z-1)/\rho$ for $x$, then we only need to show

$$Q''(x)\left(x+\frac{1}{\rho}\right) + Q'(x) \geq 0, \quad (112)$$

for $x > 0$, and (112) becomes

$$\frac{1}{\{F(x)\}^3}\left[\left[-f'(x)F(x) + 2\{f(x)\}^2\right]\left(x+\frac{1}{\rho}\right) - f(x)F(x)\right]$$
$$\geq \frac{1}{\{F(x)\}^3}\left[\left[-f'(x)F(x) + 2\{f(x)\}^2\right]x - f(x)F(x)\right], \quad (113)$$

because $-f'(x)F(x) + 2\{f(x)\}^2 \geq 0$ by (103). We claim

$$\left[-f'(x)F(x) + 2\{f(x)\}^2\right]x - f(x)F(x) \geq 0. \quad (114)$$

This is because

$$e^{-2x}\frac{x^{n-1}}{(n-1)!}\left[(x-n)\left(e^x - \sum_{k=0}^{n-1}\frac{x^k}{k!}\right) + \frac{x^n}{(n-1)!}\right] \geq 0. \quad (115)$$

Thus $-1/F_Z(z)$ is concave. By the convex ordering with the cdf $F_Y(y) = 1 - e^{-y}$ $(0 < y < \infty)$,

$$F_Z(\mathbb{E}[Z_{(m)}]) = F\left(\frac{2^{\mathbb{E}[Z_{(m)}]}-1}{\rho}\right) \geq 1 - \frac{1}{m} \quad (116)$$

$$\implies \frac{2^{\mathbb{E}[Z_{(m)}]}-1}{\rho} \geq F^{-1}\left(1-\frac{1}{m}\right) = q_m, \quad (117)$$

and therefore

$$\bar{C}(\rho) = \mathbb{E}[Z_{(m)}] \geq \log_2(1+\rho \cdot q_m). \quad (118)$$

∎

## References

[1] I. E. Telatar, "Capacity of multi-antenna Gaussian channels," *European Transactions on Telecommunications*, vol. 10, no. 6, pp. 585–595, Nov. 1999.

[2] M. Z. Win and J. H. Winters, "Analysis of Hybrid Selection/Maximal-Ratio Combing in Rayleigh Fading," *IEEE Transactions on Communications*, vol. 47, no. 12, pp. 1773–1776, Dec. 1994.

[3] R. S. Blum and J. H. Winters, "On Optimal MIMO With Antenna Selection," *IEEE Communications Letters*, vol. 6, no. 8, pp. 322–324, Aug. 2002.

[4] S. Sanayei and A. Nosratinia, "Asymptotic Capacity Analysis of Transmit Antenna Selection," in *Proc. of International Symposium on Information Theory*, p. 242, June–July 2002.

[5] N. Sharma and L. H. Qzarow, "A Study of Opportunism for Multiple-Antenna Systems," *IEEE Transactions on Information Theory*, vol. 51, no. 5, pp. 1808–1814, May 2005.

[6] A. F. Molisch, M. Z. Win, Y. S. Choi, and J. H. Winters, "Capacity of MIMO Systems With Antenna Selection," *IEEE Transactions on Wireless Communications*, vol. 4, no. 4, pp. 1759–1772, July 2005.

[7] Z. Chen, J. Yuan, and B. Vucetic, "Analysis of Transmit Antenna Selection/Maximal-Ratio Combining in Rayleigh Fading Channels," *IEEE Transactions on Vehicular Technology*, vol. 54, no. 4, pp. 1312–1321, July 2005.

[8] B. M. Hochwald, T. L. Marzetta, and V. Tarokh, "Multi-antenna channel hardening and its implication for rate feedback and scheduling," *IEEE Transactions on Information Theory*, vol. 50, no. 9 , pp. 1893–1909, Sept. 2004.

[9] H. A. David and H. N. Nagaraja, *Order Statistics*, 3rd ed., New Jeresy: John Wiley & Sons, 2003.

[10] J. Galambos, *The Asymptotic Theory of Extreme Order Statistics*, 2nd ed., Florida: Robert E. Krieger, 1987.

[11] W. R. van Zwet, *Convex Transformations of Random Variables*, Amsterdam: Mathematical Centre Tracts, 1964.

[12] P. Hall, "On the rate of convergence of normal extremes," *Journal of Applied Probability*, vol. 16, no. 2, pp. 433–439, June 1979.